\documentstyle[seceq,epsf,wrapft]{ptptex}

\title{Comparative Analysis of the Mechanisms of Fast Light Particle Formation in
Nucleus-Nucleus Collisions at Low\\ and Intermediate Energies}

\author {A.S. {\sc Denikin}\footnote {E-mail: denikin@jinr.ru}
~and V.I. {\sc Zagrebaev}}
\inst{Joint Institute for Nuclear Research, Dubna, 141980 Russia}

\recdate{February 5, 2001; in final form, May 25, 2001}

\abst
{The dynamics and the mechanisms of preequilibrium-light-particle formation in
nucleus-nucleus collisions at low and intermediate energies are studied on the basis of a
classical four-body model. The angular and energy distributions of light particles from
such processes are calculated. It is found that, at energies below 50 MeV per nucleon,
the hardest section of the energy spectrum is formed owing to the acceleration of light
particles from the target by the mean field of the projectile nucleus. Good agreement
with available experimental data is obtained.}

\begin{document}
\maketitle

\vskip -9.09cm Physics~of~Atomic~Nuclei (transl. from Yad. Fiz.)\ {\bf65} (2002), 1459 --
1473 \vskip 9.09cm

\section{INTRODUCTION}
The formation of preequilibrium light particles ($n$, $p$, $d$, $t$, $\alpha$) in
nucleus-nucleus collisions is determined by the evolution that the nuclear system
involved in the reaction being considered undergoes at the initial reaction stage. It is
well known that the cross section for light-particle yield from heavy-ion collisions
constitutes a significant part of the total reaction cross section even at low energies
of about 10 MeV per nucleon; that is, the formation of light particles is peculiar to all
nuclear reactions featuring heavy ions. This implies that investigation of the mechanism
of formation of such particles may furnish direct information both about the dynamics of
the initial stage of the reaction and about the potential and dissipative forces of
nucleus-nucleus interaction. Basic regularities in the behavior of the angular and energy
distributions of light particles - in particular, the presence of high-energy components
in them - cannot be described within the usual statistical model of excited-nucleus decay
\cite{r1,r2,r3,r4,r5,r6,r7,r8,r9,r10,r11,r12,r13,r14,r15,r16}. A large number of
theoretical approaches considering various mechanisms of fast-light-particle formation
have been proposed in recent years. These include the moving-source model \cite{r3}, the
hot-spot model \cite{r17}, the model of disintegration and incomplete fusion
\cite{r18,r19,r20,r21}, the model of dissipative disintegration accompanied by the
massive-transfer process \cite{r22,r23}, and the Fermi-jet model and models close to it
in spirit \cite{r24,r25,r26,r27,r28,r29}. A detailed survey of experimental and
theoretical studies devoted to this problem can be found in \cite{r30}. In view of a
considerable improvement of the technical characteristics of measuring equipment, it
became possible to measure precisely the angular and energy spectra of light particles.
The most recent experiments discovered preequilibrium light particles whose velocities
are more than twice as great as the velocity of beam particles \cite{r11,r12,r13,r14}.
This sparked anew the interest of researchers in the problem and reinforced motivations
behind the hypothesis that nucleon-nucleon collisions play a dominant role in the
formation of ultrafast light particles. In the present study, the role of nucleon-nucleon
collisions and of mean nuclear fields in the formation of the spectra of preequilibrium
light particles is investigated in detail on the basis of the four-dimensional classical
model of nucleus-nucleus collisions. Among other things, it is shown that the effect of
mean nuclear fields is crucial at beam energies in the region $E_0 < 50$ MeV per nucleon.
The ensuing exposition is organized as follows. In Section 2, we give an account of the
model that underlies the present analysis of the methods used here to calculate the
differential cross sections for light particles formed in nucleus-nucleus collisions and
the multiplicities of these particles. In Section 3, we consider various mechanisms of
preequilibrium-light-particle formation that are realized in the model developed here. In
Section 4, the results of our calculations for the above cross sections are compared with
experimental data. In Section 5, we investigate the dependence of our results on physical
model parameters, such as potentials of fragment interaction and forces of nuclear
friction. In the last section, we formulate basic conclusions that can be drawn from our
study.

\section{FOUR-BODY MODEL OF NUCLEUS-NUCLEUS COLLISIONS}

In studying the mechanisms of light-particle formation in nucleus-nucleus collisions, we
rely here on a semi-classical four-body model that makes it possible to establish
explicitly the role of mean nuclear fields and the role of nucleon-nucleon collisions.
Within this model, the projectile ($P$) and the target ($T$) nucleus are taken in the
form of two-particle subsystems; that is, $P = (A + a)$ and $T = (B + b)$, where $A$ and
$B$ stand for heavy nuclear cores, while $a$ and $b$ represent light fragments ($n$, $p$,
$d$, $t$, $\alpha$). Introducing six pair interaction potentials $V(r_{ij})$ (where the
subscripts $i$ and $j$ correspond to particles $A$, $B$, $a$, and $b$ and where
$\vec{r}_{ij} = (\vec {r}_i - \vec{r}_j)$ is the distance between particles $i$ and $j$),
we specify the Hamiltonian for the system being considered as
\begin{equation} H=\sum\limits_i\frac{\vec{p}_i^{\hskip 3pt 2}}{2m_i} + \sum_{ij,i \ne j} V(r_{ij}).
\label{eq1}\end{equation}
The potentials taken to represent the heavy-core interaction with light particles are
chosen here in the Woods-Saxon form with the parameters corresponding to optical
potentials constructed on the basis of an analysis of elastic-scattering data \cite{r31}.
The interaction between cores $A$ and $B$ is chosen in the form of the Coulomb potential
energy and the nuclear interaction simulated by either the proximity potential from
\cite{r32} or the Woods-Saxon potential. Coupling to reaction channels that are not taken
explicitly into account within the four-body model was described in terms of dissipative
forces introduced with the aid of the corresponding dissipative function $D$. In order to
solve numerically the set of equations of motion
\begin{equation} \frac{d\vec{r}_i}{dt}=\frac{\partial H}{\partial \vec{p}_i};\hskip 10pt
\frac{d\vec{p}_i}{dt} = -\frac{\partial H}{\partial \vec{r}_i} - \frac{\partial
D}{\partial \vec{v}_i} \label{eq2}\end{equation}
where $p_i$ and $v_j$ are the vectors of, respectively, the momentum and the velocity of
particle $i$, it is necessary to preset boundary conditions for the vectors $\vec{r}_i$
and $\vec{p}_i$. The internal spatial configuration of the projectile nucleus is
completely determined by the vector $\vec{r}_{Aa}$ of the relative distance between the
projectile fragments, the energy $E_{Aa}$ of their relative motion (that is, the
projectile binding energy), and the vector $\vec{l}_{Aa}$ of the orbital angular momentum
associated with the relative motion of these particles. The components of the vector
$\vec{r}_{Aa}(t = 0)$ are chosen via their generation at random on the basis of some
spatial-distribution function. Our calculations revealed that the form of the radial
dependence of this distribution affects only slightly the final result. This is explained
by the specificity of the classical model, where, as the time of approach of the nuclei
involved increases, any initial distribution tends to a purely classical distribution, in
which case the particle resides for a longer time in the vicinity of the external turning
point. At the same time, a decrease in the time of approach entails an increase in the
computational error. In the case being considered, the relative position of particles $A$
and $a$ was chosen to be equiprobable in the energetically allowed region of space. In
order to determine the relative momentum $\vec{p}_{Aa}$ unambiguously, it is necessary to
fix, in addition to the relative energy $E_{Aa}$, the distance $\vert\vec{r}_{Aa}\vert$
between the fragments, and the orbital angular momentum $\vert\vec{l}_{Aa}\vert$, one of
the components of the vector $\vec{l}_{Aa}$ as well (this is also done via a generation
at random). Applying the same procedure to the target nucleus and specifying the relative
motion of the centers of mass of the target and the projectile nucleus in accordance with
a given reaction, we fully define boundary conditions that are necessary for solving the
set of Eqs. (\ref{eq2}). The function $D$ in Eqs. (\ref{eq2}) is an ordinary Rayleigh
dissipative function that describes the dissipation of energy and of the angular momentum
of the relative motion of the nuclei involved. In the case where fragments $a$ and $b$
are much lighter than cores $A$ and $B$, it is assumed that the friction forces act only
between the cores. In terms of spherical coordinates, the Rayleigh function then has the
diagonal form
\begin{eqnarray}& D = \frac{1}{2} f(r)\left(\gamma_r \dot r^2 + \gamma_\theta r^2 \dot\theta^2 + \gamma_\phi
r^2 \sin^2\theta \hskip 3pt\dot\phi^2\right) &,\\& \gamma_\theta =\gamma_\phi=\gamma_t,&
\nonumber \label{eq3}\end{eqnarray}
where $\gamma_r$ and $\gamma_t$ are, respectively, the radial and the tangential
coefficient of friction; $f(r)$ is the radial form factor for dissipative forces; and
$\vec{r} = \vec{r}_A - \vec{r}_B = \{r, \theta, \phi\}$ is the vector of the relative
motion of particles $A$ and $B$. In choosing the coefficients $\gamma_r$ and $\gamma_t$
and the form factor $f(r)$, we followed \cite{r33}. Thus, we have a set of 24 coupled
classical differential equations of motion [set of Eqs. (\ref{eq2})]; we construct here
its numerical solutions directly in the laboratory frame using Cartesian coordinates. By
performing a numerical integration of Eqs. (\ref{eq2}) with respect to time for different
initial conditions, we arrive at various output channels. It can easily be shown that,
within the four-body model used, there are 15 output reaction channels. These are
scattering channels, channels involving the breakup of the projectile or the target (or
both), particle-transfer channels, and channels of complete and incomplete fusion:
\begin{equation} P+T \equiv (Aa)+(Bb)\rightarrow \cases {(Aa)+(Bb),\cr A+a+B+b,\cr \cdots,\cr
(Ab)+(Ba),\cr (Aab)+B,\cr \cdots,\cr (ABb)+a,\cr (ABab). } \label{eq4}\end{equation}
Here, parentheses enclose bound states of two or more fragments. In describing the
relative motion of particles ($a$ + $B$) and ($b$ +$A$), the differential cross sections
for various channels can be estimated more correctly in terms of the probability of their
absorption. We define this probability $P_{ij}^{abs}$ as
\begin{eqnarray} P_{ij}^{abs} & = & 1 - \exp\left(-\frac{ s_{ij}}{\lambda_{ij}}\right) \nonumber \\
& =  & 1 - \exp \left(-\int_{tr}\frac{2W_{ij}(r^\prime)dr^\prime}{\hbar
v_{ij}(r^\prime)}\right), \label{eq5}\end{eqnarray}
where $s_{ij}$ is the distance that particle $i$ travels in nucleus $j$, $\lambda_{ij}$
is the corresponding mean range, $W_{ij}(r)$ is the imaginary part of the optical
potential (it describes absorption in the case of the elastic scattering of particle $i$
by nucleus $j$), and $v_{ij}$ is their relative velocity. Integration in (\ref{eq5}) is
performed along the actual trajectory of the fragments.

In classical dynamics, the relative energy of two particles in a bound state can take any
admissible value -- in particular, collapse onto the potential-well bottom is possible.
In studying the formation of bound states of two or more particles in output channels, it
is therefore necessary to check additionally their relative energy in order to eliminate
unphysical events where the energy of the fragments is below the experimental energy of
their bound state.

For any reaction channel in (\ref{eq4}), the differential cross section is calculated by
the formula
\begin{eqnarray}  \frac{d^2\sigma_\mu}{dEd\Omega}(E,\theta) & = & \int\limits_0^\infty2\pi\rho
\nonumber \\ & & \times \left[ \frac{\Delta
N_\mu(\rho,E,\theta)}{N_{tot}(\rho)}\frac{1-P_\mu(\rho)}{2\pi sin\theta
\Delta\theta\Delta E} \right] d\rho, \label{eq6}\end{eqnarray}
where $\Delta N_{\mu}(\rho,E,\theta)$ is the number of events in which the system goes
over into the channel $\mu$, at a given value of the impact parameter $\rho$,
$N_{tot}(\rho)$ is the total number of simulated events for a given value of $\rho$, and
$P_{\mu}(\rho)$ is the probability of absorption in this channel. The bracketed factor in
the integrand on the right-hand side of (\ref{eq6}) is the partial differential
multiplicity for the event type being considered. Upon individually integrating this
factor with respect to the impact parameter, we would obtain the differential
multiplicity, which is a quantity often measured in experiments instead of the
corresponding cross section.

The main contribution to the soft section of the energy spectrum of light particles comes
from the evaporation of particles from excited reaction products. As a rule, the
multiplicity of evaporated light particles considerably exceeds the multiplicity of
preequilibrium particles in this energy range. In the proposed model, evaporation
processes are taken into account in the following way. The introduction of
phenomenological friction forces in the equations of motion (\ref{eq2}) leads to a
dissipation of part of the kinetic energy, whereby it is converted into the excitation
energy of heavy fragments. Since the problem of how the excitation energy is shared among
colliding nuclei has not yet been solved conclusively, we describe here the evaporation
section of the spectrum, assuming that the total excitation energy is shared among
colliding nuclei according to the simplest mechanism of equality of their temperatures,
in which case the excitation energy is shared in proportion to the masses of the
colliding nuclei. In the source rest frame, evaporated particles have a Maxwell
distribution with respect to energy and an isotropic angular distribution. In the
laboratory frame, the distribution of light fragments evaporated from the $i$th source
has the form
\begin{eqnarray} \lefteqn{f_i(\rho,E,\theta,\phi) = \frac{1}{2(\pi T_i(\rho))^{3/2}} \sqrt{E-V_C} }
\nonumber \\ & \times \exp\left( -{\frac{E-V_C+\varepsilon_i(\rho) - 2\sqrt{(E-V_C)\hskip
3pt \varepsilon_i(\rho)} \cos\theta^\prime}{T_i(\rho)}}\right). \label{eq7}\end{eqnarray}
Here, $E$ is the laboratory energy of the light particle; $V_{C}$ is the height of the
Coulomb barrier for this particle in escaping from a heavy fragment; $\varepsilon_i(\rho)
= mv_i^2(\rho)/2$, where $m$ is the mass of the emitted light particle and $v_{i}$ is the
laboratory velocity of the $i$th evaporative heavy fragment; $T_i(\rho) =
\sqrt{E^*_i(\rho)/a_i}$ is its temperature, where $E^*(\rho)$ is the fragment excitation
energy; $a_i$ is the level-density parameter in the corresponding nucleus; and
$\cos\theta^\prime = \sin\theta \sin\theta_i (\cos\phi \cos\phi_i + \sin\phi \sin\phi_i)
+ \cos\theta \cos\theta_i$, with ($\theta_i, \phi_i$) and ($\theta, \phi$) being the
spherical angles of emission of, respectively, the $i$th hot fragment and the evaporated
particle in the laboratory fragment. The quantities $\varepsilon_i$ and $T_i$, as well as
the angles $\theta_i$ and $\phi_i$, are functions of the impact parameter $\rho$ and are
calculated by performing averaging over the total number of events at given $\rho$ that
have resulted in the formation of the $i$th fragment. The averaging of the function
$f_i(\rho,E,\theta,\phi,\theta_i,\phi_i)$ over the azimuthal angle $\phi_i$ can be
performed analytically, whereupon the dependence on the light-particle emission angle
$\phi$ also disappears. In general, averaging over the polar angle $\theta_i$ of the
heavy fragment can be performed only numerically. Within the model used here, three types
of evaporated fragments can be formed. These are a projectile-like fragment (PLF), a
targetlike fragment (TLF), and a compound nucleus (CN). In general, we therefore obtain
three evaporation components of the energy spectrum of light particles. Within the model
proposed here, the double-differential cross section for the formation of evaporated
light particles is calculated by the formula
\begin{eqnarray} \frac{d^2\sigma^{EV}}{dEd\Omega}(E,\theta)=\int\limits_0^{\rho_{max}}2\pi\rho [
P_{CN}(\rho)C_{CN}f_{CN} \nonumber \\ + (1-P_{CN}(\rho))(C_{PLF}f_{PLF}+C_{TLF}f_{TLF})]
d\rho, \label{eq8}\end{eqnarray}
where $P_{CN}(\rho)=\Delta N_{CN}(\rho)/N_{tot}(\rho)$ is the probability of the
formation of a compound nucleus in a collision occurring at an impact parameter $\rho$
and $C_i$ are constant normalization factors. These factors were introduced in order to
normalize correctly the evaporation spectrum to experimental data; as a matter of fact,
they are proportional to the measured value of the multiplicity of evaporated particles.
The experimental normalization of the evaporation section of the spectrum of light
particles makes it possible to single out their relative contribution to the total cross
section, whereupon we can focus on preequilibrium light particles, which are the subject
of our main interest.

\section{MECHANISMS OF PREEQUILIBRIUM LIGHT PARTICLE FORMATION}
Within the proposed model, there are eight reaction channels contributing to the total
cross section for the formation of preequilibrium light particles -- fragments $a$ or/and
$b$. There are
\begin{equation} (Aa)+(Bb)\rightarrow \cases {\begin{array}{ll}A+a+B+b,\cr (Bb)+A+a,\cr
(Aa)+B+b,\end{array}\cr \begin{array}{ll}(Ab)+B+a,\cr (Ba)+A+b,\end{array}\cr
\begin{array}{ll}(AB)+a+b,\cr (ABb)+a,\cr (ABa)+b.\end{array}\cr } \label{eq9}\end{equation}
Our calculations revealed that, even at beam energies of about 30 MeV per nucleon, the
main contribution to the cross section for light-particle formation comes from the
breakup and breakup-transfer channels. Channels featuring a bound state of the heavy
cores ($ABx$) contribute significantly only at low energies ($E_0 < 20$ MeV per nucleon).

From the scheme given by (\ref{eq9}), it can be seen that the set of preequilibrium light
particles can be broken down into two subsets including particles emitted from the
projectile (particle $a$) and particles emitted from the target (particle $b$). Thus,
three evaporation components of the energy spectrum of light particles are supplemented
with two components of preequilibrium light particles. In just the same way as in the
case of the evaporation spectra of light particles, it is necessary to introduce constant
normalization factors for the preequilibrium target and the preequilibrium projectile
component, since, in the four-body model used here, the multiplicity is always less than
or equal to two. The values of these factors were chosen in such a way that the
calculated cross sections at the tails of the energy distributions would coincide in
amplitude with experimental cross sections, since, in this region of the spectrum, only
preequilibrium particles contribute.

\begin{figure}
\epsfxsize = 60 mm \centerline {\epsfbox{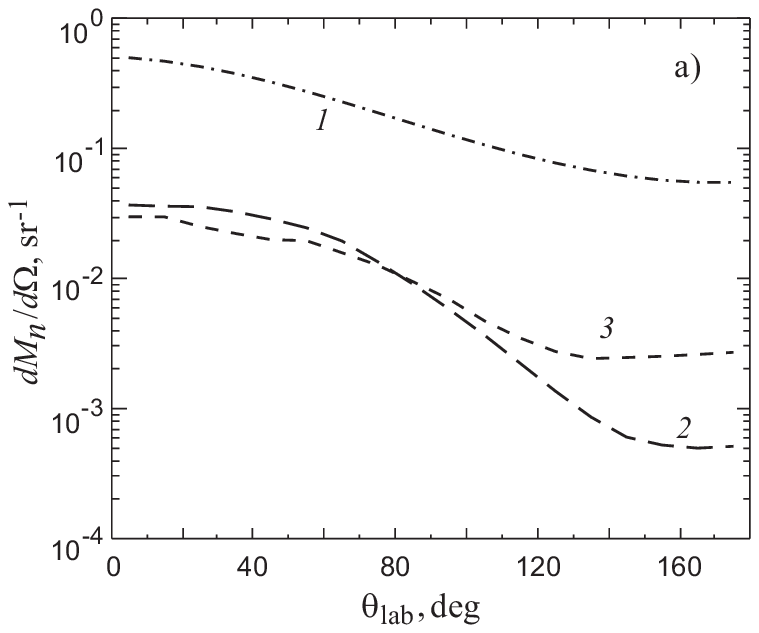} \epsfxsize = 60 mm
\epsfbox{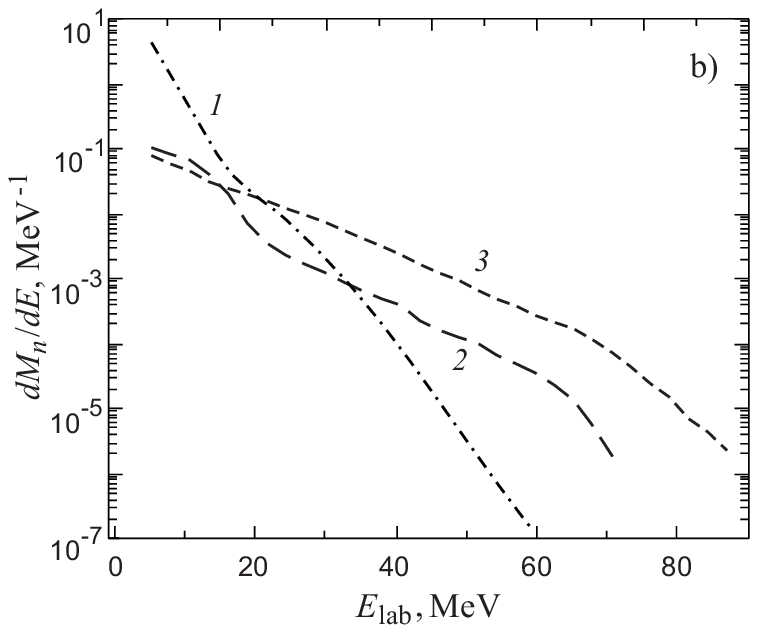}} \caption{Differential multiplicity of neutrons produced in
${}^{20}Ne$(20 A MeV) + ${}^{165}Ho \to n + X$ reactions versus (a) the neutron emission
angles and (b) the neutron energy. Curves 1 represent the total contribution of
evaporated neutrons, while curves 2 and 3 depict the contributions of preequilibrium
neutrons emitted the projectile and the target, respectively.} \label{fig:1}
\end{figure}

As a first example, we applied the proposed model to studying the properties of neutrons
emitted in $^{20}Ne + {}^{165}Ho \to n + X$ reactions at a beam energy of $E_0$ = 20 MeV
per nucleon. In this case, the target and the projectile nucleus are both represented as
a bound state of a core and a neutron: $^{20}Ne = {}^{19}Ne + n$ and $^{165}Ho =
{}^{164}Ho + n$. The binding energies of these systems were chosen on the basis of
experimental data. Figure \ref{fig:1} shows the results of our calculations for the (a)
angular and (b) energy distributions of neutrons for the above reaction. Curve 1
corresponds to equilibrium neutrons evaporated from PLF, TLF, and CN fragments of the
reactions. Curves 2 and 3 represent the contributions of preequilibrium neutrons emitted
from the projectile (particle $a$) and from the target (particle $b$), respectively.

In plotting the angular distribution displayed in Fig. \ref{fig:1}a, integration of the
differential multiplicity $d^2M_n/(dEd\Omega)$ with respect to energy was performed with
the energy spectrum cut off in the low-energy section ($E_n > 5$ MeV). It can be seen
that the evaporation component is dominant over the entire angular range, both
preequilibrium components being forward directed to a considerable extent.

As can be seen from Fig. \ref{fig:1}b, it is the preequilibrium components (curves 2 and
3) that make a dominant contribution to the energy distribution at high neutron energies
(at their velocities higher than the velocity of beam particles). That the hardest part
of the spectrum corresponds to neutrons emitted from the target nucleus (and not from the
projectile nucleus, as has usually been assumed so far) is a remarkable fact, which could
not be anticipated from the outset. There is a simple explanation of this phenomenon,
which is quite unusual at first glance.

\begin{figure}
\epsfxsize = 60 mm \centerline {\epsfbox{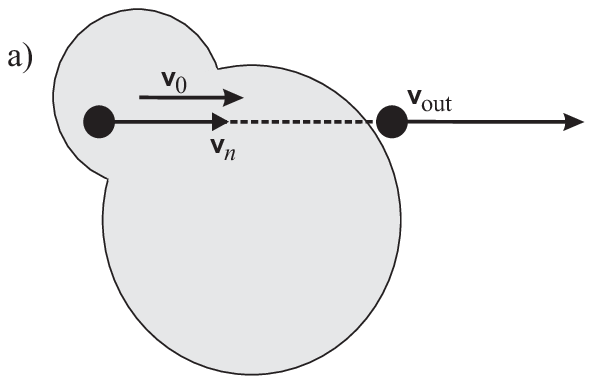} \epsfbox{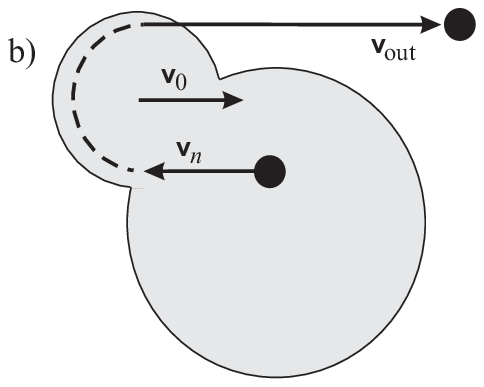}}
\end{figure}
\begin{figure}
\epsfxsize = 60 mm \centerline {\epsfbox{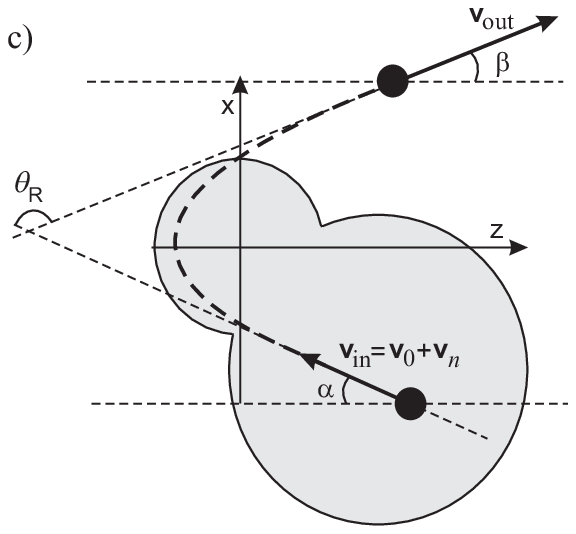}} \caption{Schematic representation of
preequilibrium nucleon formation in a nucleus-nucleus collision: (a) emission of a
nucleon from the projectile, (b) acceleration of a nucleon under the conditions of
orbiting in the projectile field (this is equivalent to scattering on a moving reflecting
wall), and (c) general case of nucleon emission from the target in projectile rest frame
(for the notation, see main body of the text).} \label{fig:2}
\end{figure}

Under the assumption that the core masses are much greater than the neutron mass, we will
now calculate the maximum possible values that kinematics allows for the energy of
neutrons emitted from the projectile and from the target. In doing this, we disregard the
neutron-neutron interaction and the distortion of the trajectories of the heavy fragments
$A$ and $B$. The velocity of the projectile neutron in the laboratory frame (see Fig.
\ref{fig:2}a) is equal to the sum of the beam-particle velocity $\vec{v}_0$ ($E_0 = m_n
v^2_0/2$) and the velocity of the internal motion of the neutron in the mean field
$\vec{v}_n$ of the projectile nucleus (in the case of a square well, we have $m v_n^2/2 -
U_0 = -E_{sep}$). Thus, the projectile nucleon emitted from a nucleus-nucleus collision
has the energy
\begin{eqnarray} E_n & = &\left( \frac{m}{2}(\vec{v}_0+\vec{v}_n)^2-U_0\right)
\bigl\vert_{v_n=v_F,\theta_n=0^\circ} \nonumber \\ & = & \frac{m}{2} \left( v_0^2
+2v_0v_F\right)-E_{sep}, \label{eq10}\end{eqnarray}
where $m$ is the neutron mass, $U_0$ is the depth of the mean field $V_{Aa}$, $E_{sep}$
is the neutron-separation energy, and $v_F$ is the Fermi velocity of projectile nucleons.
If, for example, $v_0 \sim v_F$ and $E_0 \gg E_{sep}$, the maximum energy of the emitted
nucleon is $E^{max}_n \sim 3E_0$. This mechanism of fast-light-particle formation was
comprehensively investigated in \cite{r25,r29}.

The mechanism responsible for the formation of high-energy neutrons from the target is
more complicated. The main role in this process is played by the potential of the
interaction between the projectile nucleus $A$ and the target neutron $b$. Let us first
consider a simplified model where the interaction $V_{Ab}$ is replaced by the interaction
of neutron $b$ with an infinitely heavy moving wall. Suppose that $b$ moves at a velocity
$v_n$ toward the core $A$, which, in turn, has a velocity $v_0$ directed oppositely. In
their c.m. frame, the neutron velocity is $(v_0 + v_n$); after an elastic collision, the
neutron acquires the velocity $-(v_0 + v_n)$, which corresponds to the velocity $v_{out}
= (2v_0 + v_n)$ in the laboratory frame. Under the condition that the internal-motion
velocity of the target neutron, $v_n$, is equal to the Fermi velocity $v_F$, its
asymptotic energy is
\begin{eqnarray}
E_n\vert_{v_n=v_F,\theta_n=0^\circ} & = & \frac{m}{2}(2v_0 + v_F)^2 -U_0 \nonumber \\ & =
& 2m(v_0^2 + v_0v_F)-E_{sep}, \label{eq11}\end{eqnarray}
where $U_0$ is the depth of the mean field $V_{Bb}$ and
is the energy of target-neutron separation. At $v_0 \sim v_F$ and $E_0 \gg E_{sep}$, the
maximum energy is thus $E_n^{max} \sim 8E_0$, which is 2.5 times as great as the
corresponding limit for neutrons emitted from the projectile.

The elastic scattering of neutrons at an angle of $\theta_{cm} = -180^\circ$ in the
attractive mean field of the projectile nucleus is kinematically equivalent to their
reflection from a repulsive wall (see Fig. \ref{fig:2}b). The scattering of neutrons at
such large angles ($\theta_{cm} < -180^\circ$), which is actually an orbiting process, is
possible only at comparatively low neutron energies. At c.m. energies of $E_0 \geq A E_F
\sim$ 40 A MeV, neutrons can be deflected by the mean field ($U_0 \sim 50$ MeV) by not
more than at a limiting negative angle $\theta_R$ that is referred to as the
rainbow-scattering angle. By virtue of this, the maximum energy of neutrons emitted from
the target depends strongly on the projectile energy, on the interaction potential
$V_{Ab}$, on the neutron binding energy in the target, and on the friction forces.
Disregarding the effect of neutrons on the motion of heavy fragments, assuming that the
neutron acquires the maximum energy upon scattering by the projectile at the angle
$\theta_R$ in the neutron- projectile c.m. frame (see Fig. \ref{fig:2}c), and setting the
initial neutron velocity in the target to the relevant Fermi velocity, we can estimate
the asymptotic neutron energy (at $v_n = v_F$) as
\begin{eqnarray} E_n & = & m \biggl( v_0^{2^{}} + v_0v_F\cos\alpha \biggr. \nonumber \\ & & \biggl. + v_0 \cos\beta
\sqrt{v_0^2+v_F^2+2v_0v_F\cos\alpha}\biggr)-E_{sep} \label{eq12}\end{eqnarray}
where $\alpha$ is the angle at which the neutron is incident on the target in the
reference frame comoving with the target and $\beta$ is the emission angle in the same
reference frame (see Fig. \ref{fig:2}c). The two angles are related to each other through
the nuclear-rainbow-scattering angle, for which there is the empirical relation
\cite{r34}
\begin{equation} \theta_R  = \left(V_C - 0.56U_0\sqrt{R_V/a_V}\right) \Bigl/ E_{cm}\Bigr.
= - \bigl|~\pi -\alpha - \beta~\bigr|, \label{eq13} \end{equation}
where $V_C$ is the height of the Coulomb barrier (it is equal to zero for a neutron),
while $U_0$, $R_V$, and $a_V$ are, respectively, the depth, the range, and the
diffuseness of the interaction potential $V_{Ab}$. It turned out that the empirical
formula (\ref{eq13}), with the coefficient 0.56, agrees poorly with the exact classical
calculation of the angle $\theta_R$ for the scattering of light particles (such as a
proton or a neutron) on nuclei; therefore, we use here the coefficient 0.7. It can be
seen that, at $\alpha = \beta = 0$ (that is, at $\theta_R = -180^\circ$), formula
(\ref{eq12}) reduces to (\ref{eq11}).

\begin{figure}
\epsfxsize = 90 mm \centerline {\epsfbox{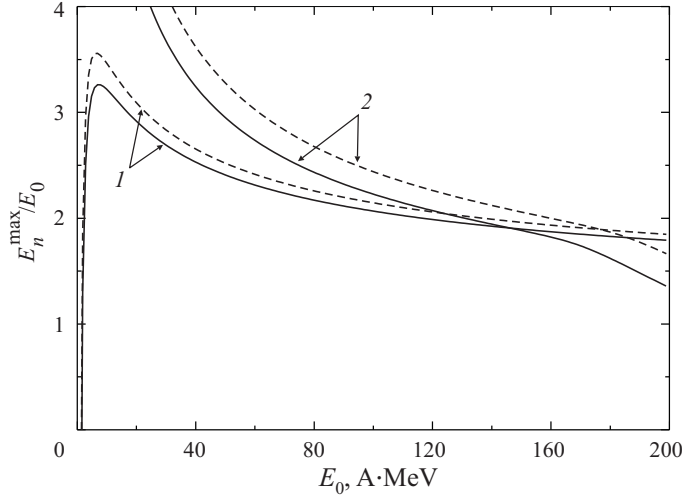}} \caption{The maximum energy
$E_n^{max}$ of preequilibrium neutrons as a function of the incident-beam energy $E_0$
for (solid curves) ${}^{20}Ne + {}^{165}Ho \to n + X$ and (dashed curves) ${}^{165}Ho +
{}^{165}Ho \to n + X$ according to the calculations based on formulas (\ref{eq10}) and
(\ref{eq12}). Curves 1 correspond to the maximum energy of neutrons emitted from the
projectile, while curves 2 represent the energies of neutrons emitted from the target. }
\label{fig:3} \end{figure}

The maximum energies of preequilibrium neutrons originating from (solid curves) $^{20}Ne
+ {}^{165}Ho$ and (dashed curves) $^{165}Ho + {}^{165}Ho$ interactions are displayed in
Fig. \ref{fig:3} versus the beam energy $E_0$. Curves 1 correspond to the results
obtained by calculating, on the basis of (\ref{eq10}), the maximum energy of neutrons
emitted from the projectile; curves 2 represent the energies of neutrons emitted from the
target, their values being calculated by formula (\ref{eq12}). From Fig. \ref{fig:3}, it
can be seen that, at energies below 100 MeV per nucleon, the fastest neutrons are emitted
from the target and that, upon going over to the heavier projectile, the maximum energy
of the emitted neutron becomes higher. This is because the range of the potential
$V_{Ab}$ increases, which entails an increase in the absolute value of the
rainbow-scattering angle $\theta_R$ [see (\ref{eq13})].

Thus, an experimental investigation of reactions where projectiles different in mass are
incident on the same target may be one of the tests of validity of conclusions that we
have drawn. For nucleons emitted from the target and the projectile to be unambiguously
identified, it is necessary that the spectrum of the projectile nucleons change
insignificantly upon going from one system to another. In this case, the distinction
between the distributions of preequilibrium nucleons will be determined completely by the
yield of precisely target nucleons. In order to ensure the invariability of the spectra
of preequilibrium nucleons emitted from the projectile, it is necessary to select
projectile nuclei with similar features (such as the angular momentum of valence nucleons
and the energies of their separation).

\begin{figure}
\epsfxsize = 90 mm \centerline {\epsfbox{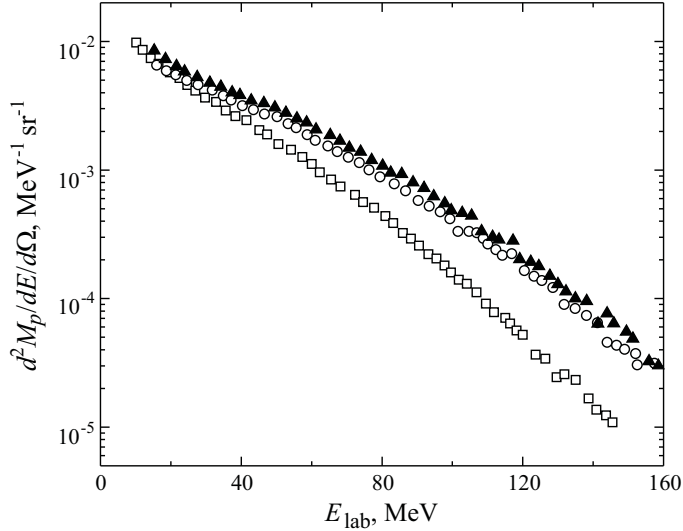}} \caption{Measured differential
multiplicities of protons emitted at an angle of $\theta_{lab} = 51^\circ$ in (open
boxes) $^{40}Ar + ^{51}V \to p + X$, (open circles) ${}^{132}Xe + {}^{51}V \to p + X$,
and (closed triangles) ${}^{132}Xe + {}^{197}Au \to p + X$ interactions at a beam energy
of 44 MeV per nucleon \cite{r16}.} \label{fig:4}
\end{figure}

Figure \ref{fig:4} shows the measured differential multiplicities of protons originating
at an angle of $\theta_{lab} = 51^\circ$ from $^{40}Ar + {}^{51}V \to p + X$, $^{132}Xe +
{}^{51}V \to p + X$, and $^{132}Xe + {}^{197}Au \to p + X$ interactions at a beam energy
of 44 MeV per nucleon \cite{r16}. It can be seen that, upon going over from the
projectile nucleus of $^{40}Ar$ to the heavier species of $^{132}Xe$, the slope of the
proton spectrum decreases, which corresponds to an increase in the yield of fast light
particles. On the contrary, the replacement of the target nucleus causes virtually no
changes in the energy distribution of protons. The change in the character of the spectra
in response to going over from one projectile-nucleus species to another can hardly be
explained by different properties of the projectile species, because the internal
structure of the projectile (the height of the Coulomb barrier, the binding energy, shell
effects, etc.) does not have a significant effect on the properties of preequilibrium
protons at the high beam energies considered here. Moreover, the mechanism of
light-particle emission from the projectile nucleus is independent of its mass-only the
multiplicity of light particles (that is, the absolute normalization of their spectrum)
depends on it. The effect of dissipative forces, which directly depend on the
target-nucleus mass, leads to a moderation of protons emitted from the projectile
nucleus. It follows that the use of a heavier target nucleus would lead to a decrease in
the yield of fast protons (because of the intensification of dissipative processes) if
they were formed only via stripping from the target. However, a comparison of the data
presented in Fig. \ref{fig:4} for $^{132}Xe$ (44 A MeV) + $^{51}V, {}^{197}Au$
interactions does not reveal any significant change in the proton spectra. Nonetheless,
the above mechanism of the acceleration of target nucleons in the mean field of the
projectile is very sensitive to the geometric dimensions of the the projectile (that is,
to its mass). Thus, we can conclude that the main contribution to the high-energy section
of the spectra displayed in Fig. \ref{fig:4} comes precisely from protons emitted from
the target and accelerated by the mean field of the projectile. The conclusion that the
energy spectrum of protons depends weakly on the choice of target nucleus also follows
from the data presented by Jasak et al. \cite{r10}, who studied the target-mass
dependence of the yields of various products (including protons) from the reactions
induced by $^{40}Ar + {}^{197}Au$ and $^{40}Ar + {}^{40}Ca$ collisions at an energy of
$E_0$ = 42 MeV per nucleon.

\begin{figure}
\epsfxsize = 90 mm \centerline {\epsfbox{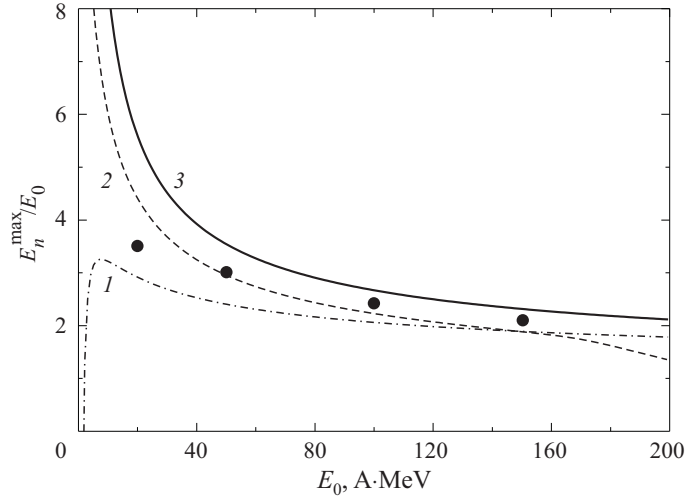}} \caption{Maximum energy of
preequilibrium neutrons originating from ${}^{20}Ne + {}^{165}Ho ->n + X$ reactions as a
function of the beam energy. Curves 1 and 2 represent the data analogous to those
depicted by the corresponding curves in Fig. \ref{fig:3}. Curve 3 corresponds to the
maximum energy acquired by the neutron upon an elastic nucleon- nucleon collision in the
mean field of the relevant dinuclear system (according to the calculation by formula
(\ref{eq14}) not allowing for the Pauli exclusion principle). The points represent the
results obtained with allowance for the Pauli exclusion principle (for details, see main
body of the text).} \label{fig:5}
\end{figure}

Within our model, we will now consider the effect of nucleon-nucleon collisions on the
formation of fast light particles. Suppose that, in the laboratory frame, a target
nucleon has a velocity $\vec{v}_j$ prior to a collision event; the velocity of a
projectile nucleon is equal to the sum of the beam-particle velocity $\vec{v}_0$ and the
nucleon velocity $\vec{v}_i$ within the projectile. One of the nucleons can acquire the
maximum velocity if, upon the collision event, it carries away the entire amount of the
relative-motion energy. In this case the maximum energy of the emitted nucleon is ($i =
a, b$)
\begin{equation}  E_i^{max} = \frac{m}{2}\left( v_b^2 + (v_0+v_a)^2\right) -\frac{mv_i^2}{2}-E_i^{sep}.
\label{eq14} \end{equation}
From (\ref{eq14}), it follows that, if $v_0 \sim v_i \sim v_F$, then $E_i^{max} \sim 4
E_0$. For $^{20}Ne + {}^{165}Ho \to n + X$ reactions, Fig. \ref{fig:5} displays the
maximum neutron energy as a function of the beam energy according to the calculations
based on formulas (\ref{eq10}) (curve 1), (\ref{eq12}) (curve 2), and (\ref{eq14}) (curve
3). It can be seen that, over the entire energy range, nucleon-nucleon collisions in this
reaction can in principle lead to the formation of yet more energetic light particles in
relation to the first two mechanisms considered above.

In actual experiments, the boundaries depicted by the curves in Fig. \ref{fig:5} will be
smeared because of the high-energy component of the momentum distributions in the
projectile and the target nucleus ($v_n > v_F$); in the case corresponding to curve 3,
there is also the contribution to this smearing from the Pauli exclusion principle, which
forbids nucleons that suffered a collision to occur in states already occupied by other
intranuclear nucleons, with the result that the probability of the acceleration of
nucleons to the maximum possible degree is considerably suppressed.

Let us introduce a nucleon-nucleon interaction featuring a repulsive core at short
distances. Solving the set of Eqs. (\ref{eq2}) for initial conditions chosen at random,
we can determine numerically the maximum energy acquired by a nucleon upon a nucleon-
nucleon collision in the mean field of the relevant dinuclear system. Testing, in the
output channels, the binding energy of the recoil nucleon, we can also take into account
the Pauli exclusion principle in our calculations. The results of our calculations for
$^{20}Ne + {}^{165}Ho \to n + X$ reactions versus $E_0$ are shown in Fig. \ref{fig:5} by
points. It can be seen that, at low initial energies ($E_0 \sim$ 20 MeV per nucleon), the
Pauli exclusion principle has a crucial effect on the formation of fast preequilibrium
neutrons in nucleon-nucleon collisions. In this case, the energy of the emitted particles
does not exceed $3.5E_0$. At higher values of the beam energy $E_0$, the discrepancy
between the predictions of formula (\ref{eq14}) and the results of the calculation
decreases gradually. It should be noted that the maximum energy of neutrons was
calculated with allowance for dissipative forces acting between the heavy cores and
exerting, as will be shown below, a pronounced effect on the spectra of preequilibrium
light particles. The dissipative forces moderate the projectile nucleus; that is, they
reduce the velocity $v_0$. This leads to an additional decrease in the quantity
$E_n^{max}$ in nucleon-nucleon collisions. With increasing initial energy $E_0$, the
effect of dissipative forces becomes less pronounced, since the valence nucleon does not
have time to "experince" the moderating influence of the projectile mean field. Thus, we
can conclude that, up to beam energies of $E_0 \sim A E_F$, the role of nucleon-nucleon
collisions is less significant than the role of the mean fields. At higher beam energies,
the maximum energy acquired by nucleons as the result of nucleon-nucleon collisions
becomes greater than the energy of target nucleons accelerated by the projectile mean
field.

\section{COMPARISON WITH EXPERIMENTAL DATA}

In order to verify the qualitative conclusions drawn in the preceding section, we have
analyzed the differential cross sections for the yield of neutrons and protons from a few
nuclear reactions and performed a comparison with available experimental data.

\begin{figure}
\epsfxsize = 65 mm \centerline {\epsfbox{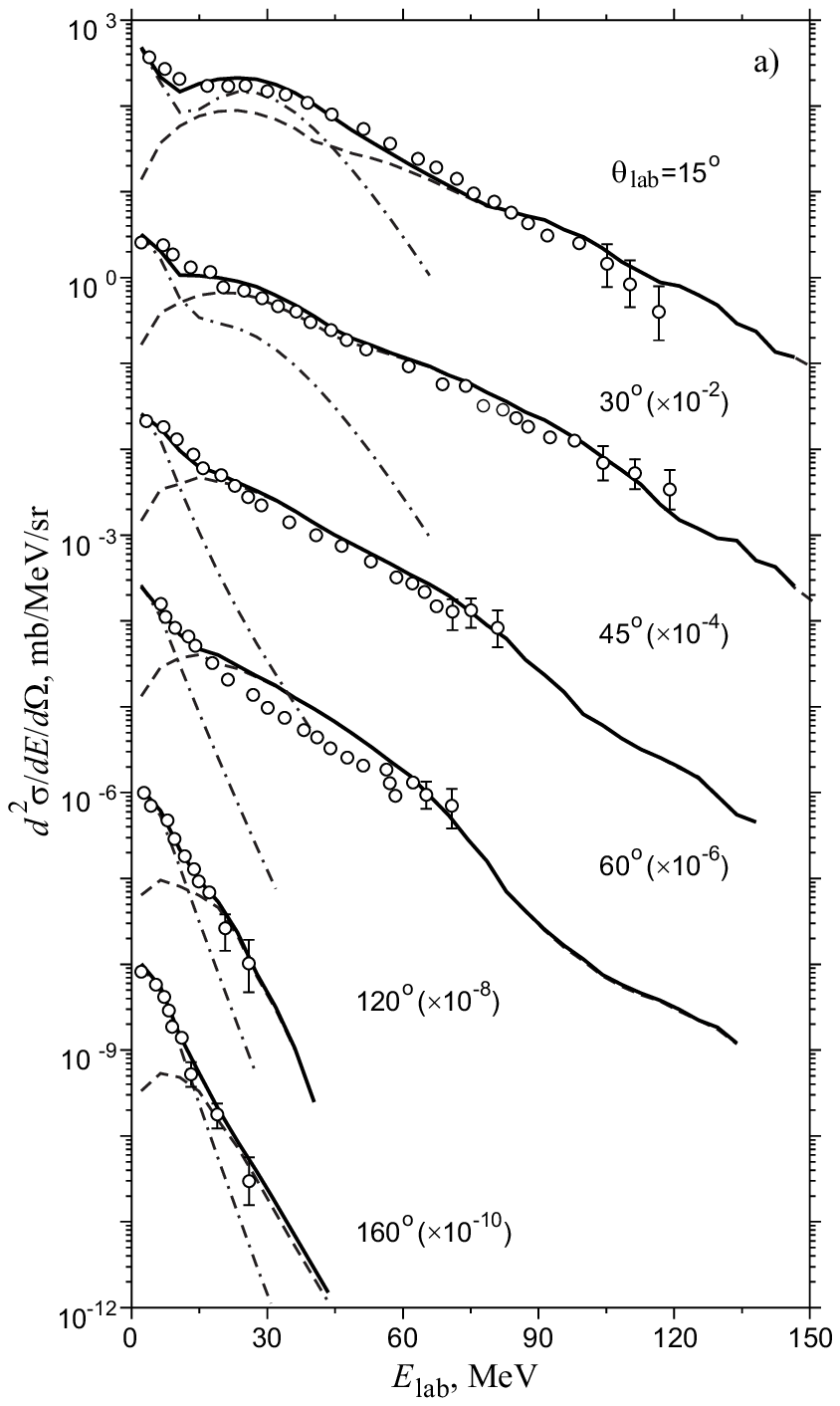} \epsfxsize = 65 mm
\epsfbox{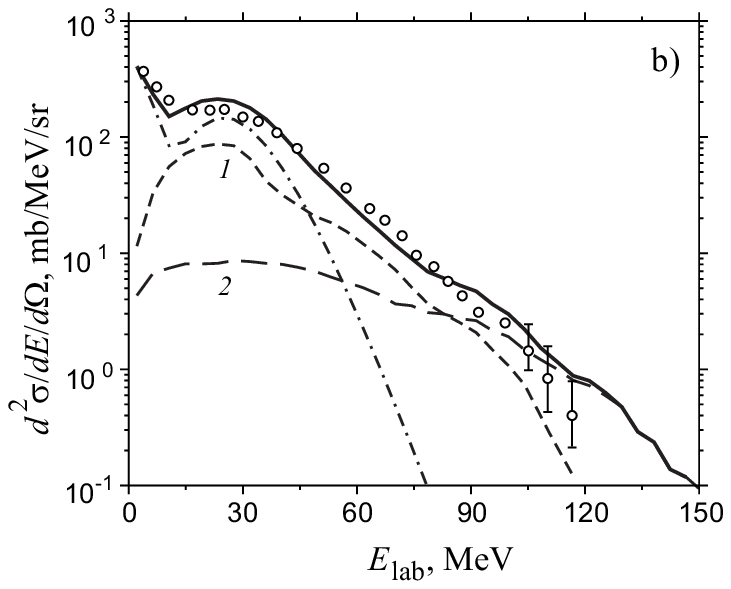}} \caption{(a) Measured and calculated differential cross sections for
neutron formation in ${}^{36}Ar$(35 A MeV) + ${}^{107}Ag -> n + X$ reactions: (points)
experimental data from \cite{r15}, (dash-dotted curves) contribution of evaporated
neutrons, (dashed curves) total contribution of preequilibrium neutrons emitted from the
projectile and the target, and (solid curves) sum of the evaporation and the
preequilibrium component; (b) theoretical results for the single angle of $\theta_{lab} =
15^\circ$ that are basically the same as in Fig. \ref{fig:6}a, except that the
contribution to the differential cross section from neutrons emitted (curve 1) by the
projectile and (curve 2) by the target are shown individually instead of their total
contribution.} \label{fig:6}
\end{figure}

The double-differential cross sections measured in \cite{r15} for the yield of neutrons
from $^{36}Ar(35 A MeV) + {}^{107}Ag \to n + X$ reactions are displayed in Fig.
\ref{fig:6}a, along with the results of the relevant calculations. The dash-dotted, the
dashed, and the solid curve represent, respectively, the evaporation component, the
preequilibrium component, and their sum. The equilibrium part of the spectrum receives
contributions from neutrons evaporated by a targetlike fragment (this is the isotropic
low-energy component completely saturating the evaporation spectrum at large angles) and
from neutrons evaporated by a projectile-like fragment, the maximum in the distribution
corresponding to forward angles and energies close to the beam energy.

Figure \ref{fig:6}b gives a more detailed pattern for the contribution of preequilibrium
neutrons emitted from (curve 1) the projectile and (curve 2) the target at an angle of
$\theta_{lab} = 15^\circ$. The solid and the dash-dotted curve are identical to their
counterparts in Fig. \ref{fig:6}a. It can be seen that the hardest section of the
spectrum is associated with neutrons emitted from the target nucleus. For intermediate
values of the emission angle ($\theta_{lab} < 90^\circ$), this trend is conserved; only
in the region of large angles are the contributions of the two preequilibrium components
approximately equal.

\begin{figure}
\epsfxsize = 65 mm \centerline {\epsfbox{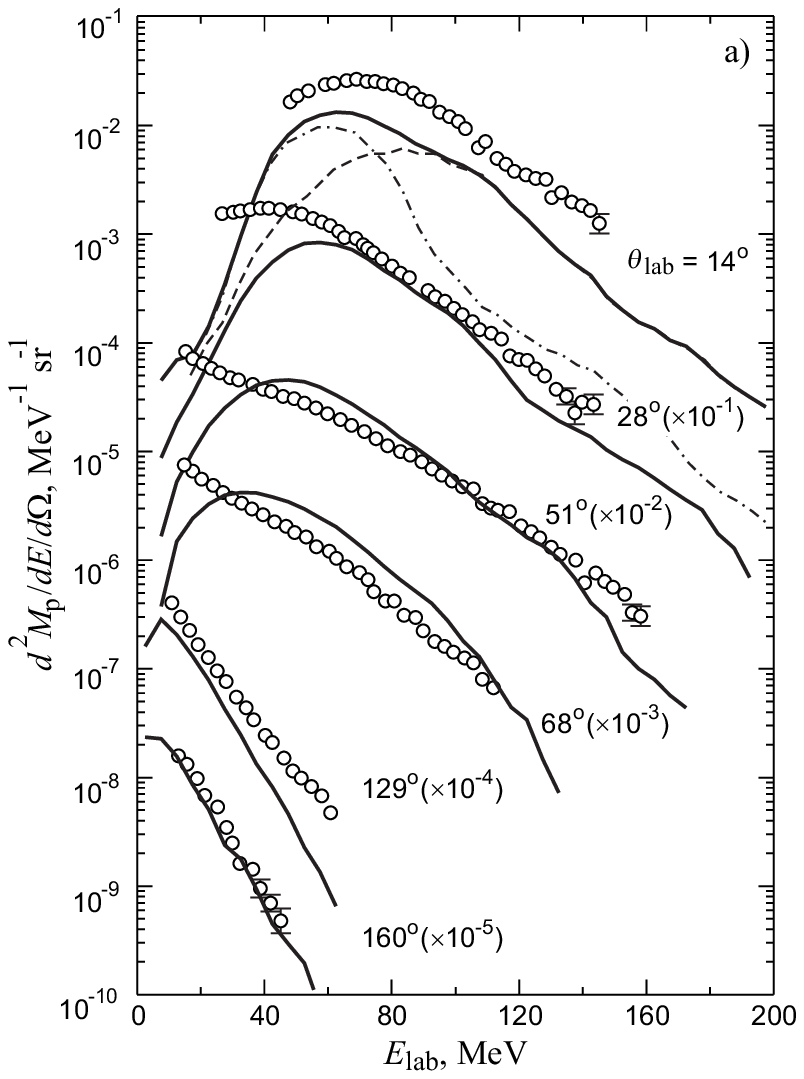} \epsfxsize = 65 mm
\epsfbox{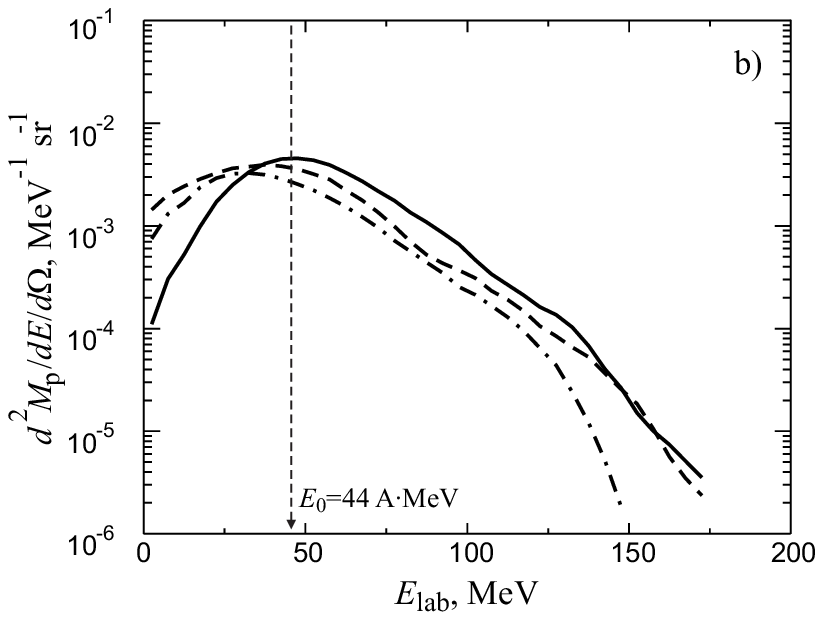}} \caption{(a) Comparison of (points) the measured \cite{r16} and
(solid curves) the calculated differential multiplicity of preequilibrium protons formed
in ${}^{132}Xe$ (44A MeV) + ${}^{197}Au -> p + X$ reactions. The dashed and the
dash-dotted curve represent the contributions of preequilibrium protons emitted at an
angle of $\theta_{lab} = 14^\circ$ from the target and the projectile, respectively. The
contribution of evaporated protons is not shown. (b) Calculated differential
multiplicities of protons emitted at an angle of $\theta_{lab} = 51^\circ$ in (solid
curve) ${}^{132}Xe + {}^{197}Au \to p + X$, (dashed curve) ${}^{132}Xe + {}^{51}V \to p +
X$, and (dash-dotted curve) ${}^{40}Ar + {}^{51}V \to p + X$ reactions at the beam energy
of 44 MeV per nucleon.} \label{fig:7}
\end{figure}

Figure \ref{fig:7}a displays experimental data from \cite{r16} on the differential
multiplicity of protons emitted in $^{132}Xe$(44 A MeV) + ${}^{197}Au \to p + X$
reactions. In that figure, the solid curves represent the calculated energy distribution
of preequilibrium protons; shown additionally for the emission angle of $\theta_{lab} =
14^\circ$ are the contributions of protons escaping from (dashed curve) the target and
(dash-dotted curve) the projectile. The contribution of evaporated protons is not
presented. The theoretical results for the proton spectrum at $\theta_{lab} = 14^\circ$
noticeably underestimate the experimental cross section in magnitude, but they reproduce
quite well the behavior of experimental data. On the contrary, the theoretical curves in
the region of backward angles ($\theta_{lab} = 129^\circ, 160^\circ$) lie somewhat above
the experimental data on the differential multiplicity of preequilibrium protons. This is
because the evaporation component must be dominant in this region. We can see that,
despite the use of quite a simple semiclassical model, the agreement with experimental
data is by and large satisfactory.

\begin{figure}
\epsfxsize = 65 mm \centerline {\epsfbox{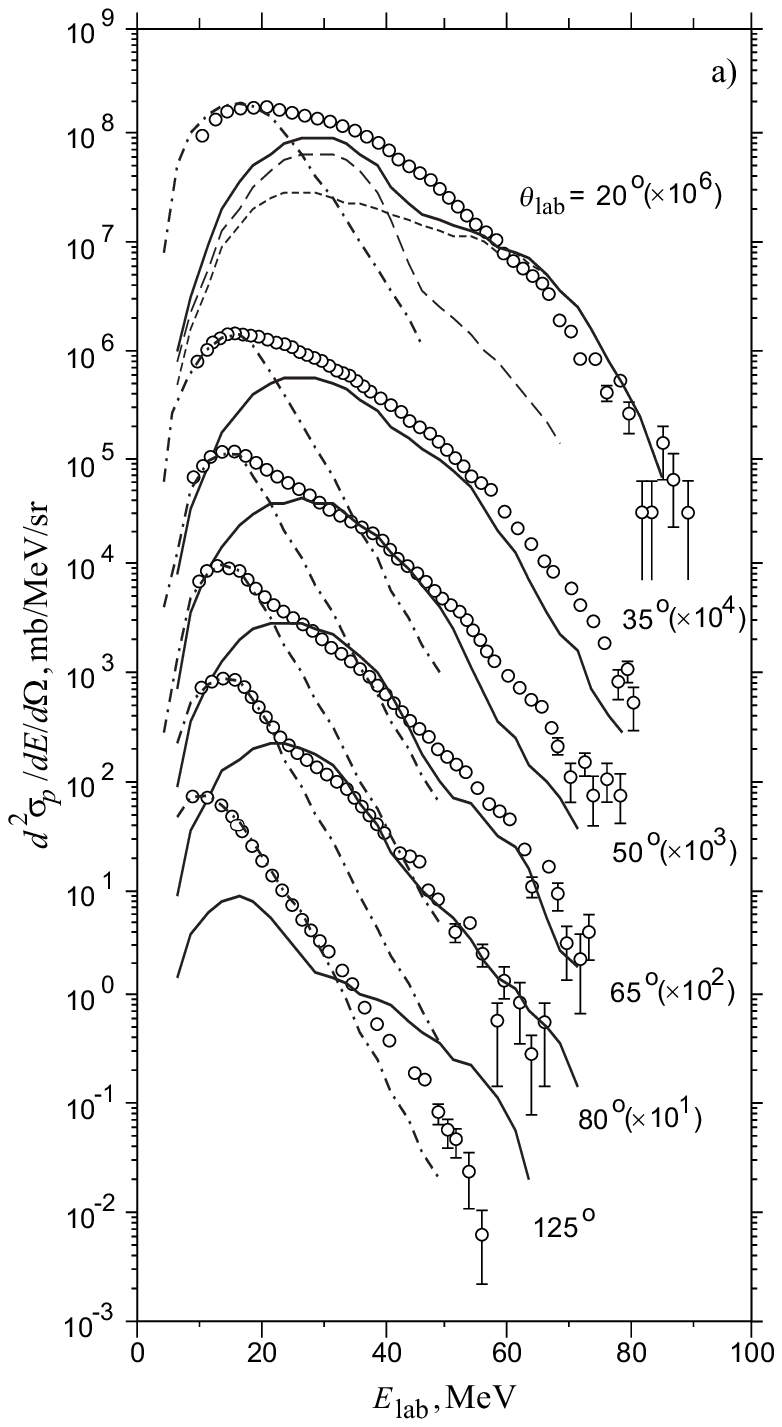} \epsfxsize = 65 mm
\epsfbox{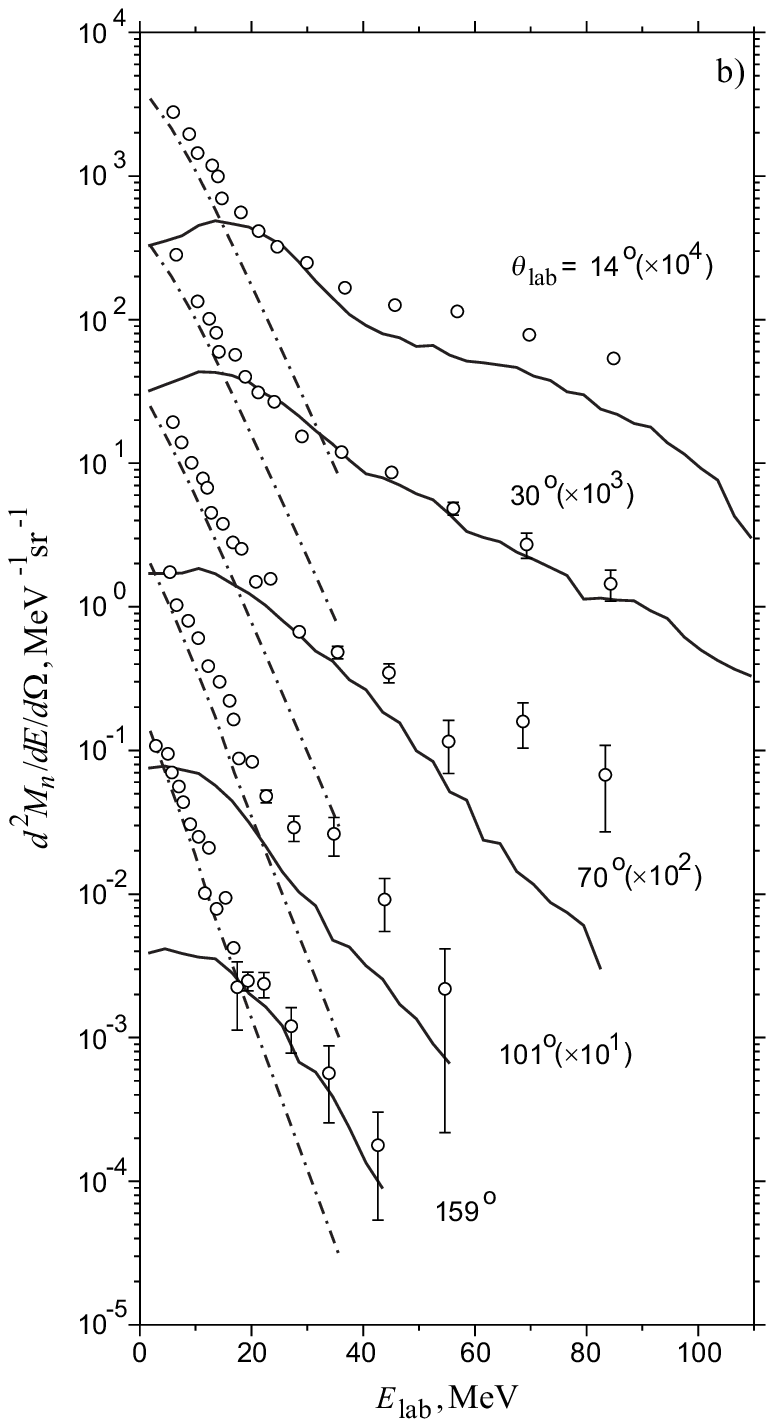}} \caption{(a) Measured\cite{r4} and calculated double-differential
cross sections for proton production in ${}^{16}O$(20 A MeV) + ${}^{197}Au \to p + X$
reactions and (b) measured\cite{r8} and calculated differential multiplicity of neutrons
from ${}^{20}Ne$(30 A MeV) + ${}^{165}Ho \to n + ER$ ($\theta_{ER} = 5.6^\circ$)
reactions: (solid curves in Figs. \ref{fig:8}a and \ref{fig:8}b) computed distributions
of preequilibrium protons and neutrons, respectively; (dash-dotted curves) contribution
to the distributions from evaporated light particles; and (points) experimental data. In
Fig. \ref{fig:8}a, the contributions of preequilibrium protons emitted from the
projectile and from the target at an angle of $\theta_{lab} = 20^\circ$ are shown
individually by long and short dashes, respectively.} \label{fig:8}
\end{figure}

Figure \ref{fig:7}b shows the computed energy distributions of preequilibrium protons
emitted at an angle of $\theta_{lab} = 51^\circ$ in (solid curve) $^{132}Xe + {}^{197}Au
\to p + X$, (dashed curve) $^{132}Xe + {}^{51}V \to p + X$, and (dash-dotted curve)
${}^{40}Ar + {}^{51}V \to p + X$ reactions at a beam energy of 44 MeV per nucleon. It can
be seen that, in the region of high energies, the spectrum of product protons is harder
for the heavier projectile of ${}^{132}Xe$ than for the lighter projectile of ${}^{40}Ar$
owing to particles emitted from the target nucleus. No such effect arises upon replacing
the target nucleus by a heavier one. Comparing the curves in Fig. \ref{fig:7}b with the
experimental data in Fig. \ref{fig:4}, we can see that the agreement between the results
of the theoretical calculations and the experimental data is quite satisfactory.

In Fig. \ref{fig:8}a, the double-differential cross section measured in \cite{r4} for
proton formation in ${}^{16}O(20 A MeV) + {}^{197}Au->p + X$ reactions is contrasted
against the results of the theoretical calculations based on the model employed here. The
solid curves correspond to the total contribution of preequilibrium protons emitted from
the target and the projectile. The dash-dotted curves represent the contribution of
evaporated protons. For protons emitted at an angle of $\theta_{lab} = 20^\circ$, more
detailed dependences are shown individually for protons escaping from (long dashes) the
projectile and (short dashes) the target. The cross sections calculated for the reactions
in question agree well with experimental data at small and intermediate values of the
emission angle. As in the preceding case, however, the cross section is overestimated at
large values of the proton emission angle.

In Ref. \cite{r8}, the differential multiplicity was measured for neutron formation in
coincidence with the evaporation residue in the reaction ${}^{20}Ne(30A MeV) + {}^{165}Ho
\to n + ER(\theta_{ER} = 5.6^\circ)$. Within the model used here, we can calculate the
differential multiplicity of neutrons in coincidence with the evaporation residue,
taking, however, no account of its emission angle $\theta_{ER}$. This limitation is due
to the fact that we can only roughly estimate the contribution of evaporation processes
to the cross section for the formation light fragments; in doing this, we underestimate
the multiplicities of preequilibrium particles, so that we cannot calculate precisely the
emission angle for the evaporation residue. In Fig. \ref{fig:8}b, the multiplicities
calculated in the present study are contrasted against the experimental data from
\cite{r8}. In that figure, the dash-dotted curve represents the total contribution of
evaporated neutrons, while the solid curve corresponds to the total distribution of
preequilibrium neutrons. At some values of the emission angle, there is a sizable
discrepancy between the results of theoretical calculations and experimental data in the
region of high energies. In all probability, this is due to imperfections of the model in
dealing with correlation experiments, which require a precise treatment of the
statistical decay of an excited nucleus. By and large, we can state that, despite the
simplicity of the proposed model, there is good agreement, in the energy range being
considered, between the calculated cross sections for the yield of light particles from
nucleus-nucleus collisions and relevant experimental data.

\section{DYNAMICS OF LIGHT-PARTICLE FORMATION AND ROLE OF DISSIPATIVE FORCES}

In treating the dynamics of nucleus-nucleus collisions on the basis of the semiclassical
approach involving two-particle interactions, there remains an ambiguity in choosing the
parameters of these interactions. In assessing the parameters of the potentials that
simulate the interaction between a light particle and a heavy fragment, we relied here on
experimental data obtained by exploring elastic scattering and on the results of their
treatment within the optical model \cite{r31}. It is well known, however, that the
optical model leads to a discrete and a continuous ambiguity in the potential parameters.
In the present calculations, these parameters were therefore varied within $10-20\%$ in
order to determine the stability of the results of these calculations and conclusions
drawn from them. In particular, the potentials $V_{Aa}$ and $V_{Bb}$ (that is, the
interaction of a valence nucleon with a nuclear core) chosen in the Woods-Saxon form had
the parameters of $U_0 = 50-60$ MeV, $r_0 = 1.15-1.25$ fm, and $a_v = 0.45-0.6$ fm.
Variations of these parameters in the above ranges do not have a strong effect on the
angular and energy distributions of preequilibrium light particles.

The potentials $V_{Ab}$ and $V_{Ba}$ play a much more significant role. As was shown
above, the former is responsible for the acceleration of the target valence nucleon and,
as a consequence, for the formation of the spectrum of preequilibrium light particles
emitted from the target. The latter determines the angular distribution of light
particles emitted by the projectile. The parameters of the potential $V_{Ab}$ specify the
rainbow-scattering angle in (\ref{eq13}), which sets a kinematical limit on the energy of
the emitted light particle b. In the calculations, the parameters of these potentials for
various nuclei were taken in the following ranges: $U_0 = 45-55$ MeV, $r_0 = 1.15-1.25$
fm, and $a_V = 0.45-0.65$ fm.

In addition to the real parts of the potentials $V_{Ab}$ and $V_{Ba}$, we also introduced
imaginary parts (see Section 2). It turned out that these imaginary parts, which
determine the absolute values of the cross sections, have virtually no effect on other
observables, such as the slope of the spectrum and the position of the maximum. In the
case being considered, the imaginary parts of the potentials $V_{Ab}$ and $V_{Ba}$, were
chosen in the Woods-Saxon form with parameters based on data from Ref. \cite{r31}.

\begin{figure}
\epsfxsize = 65 mm \centerline {\epsfbox{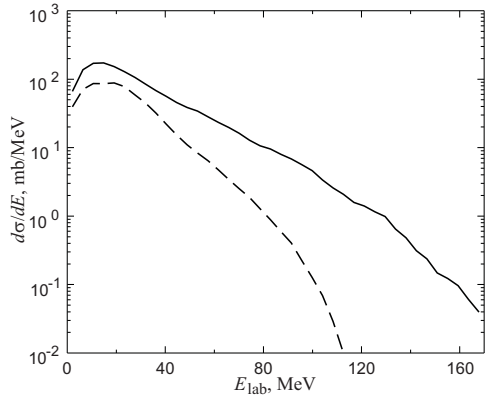}} \caption{Differential cross section
for preequilibrium neutrons originating from ${}^{36}Ar$(35 A MeV) + ${}^{107}Ag \to n +
X$ reactions for impact-parameter values from the intervals (dashed curve) $\rho \in$
[0,4] fm and (solid curve) $\rho \in$ [4,9] fm.} \label{fig:9}
\end{figure}

The potential $V_{AB}$ simulating the nucleus-nucleus interaction plays a significant
role in the formation of light particles at low beam energies of $E_0 < 20$ MeV per
nucleon, but it becomes less important as the beam energy increases. The potential
$V_{AB}$ determines the trajectories of the projectile, which is responsible for the
acceleration of target nucleons. That the projectile moves along trajectories not
coinciding with a straight line smears the region of forward angles in the spectrum of
the fastest light particles. Such particles are produced in peripheral collisions
characterized by impact-parameter values close to that of tangential collision. Figure
\ref{fig:9} shows the differential distribution of the cross section with respect to
energy for preequilibrium neutrons originating from ${}^{36}Ar$(35 A MeV) + ${}^{107}Ag$
interactions at impact-parameter values from various ranges. The dashed curve represents
the results for impact parameters satisfying the condition $\rho < 4$ fm, while the solid
curve corresponds to $\rho$ values between 4 and 9 fm. In the case of a tangential
collision, the impact parameter is about 8.5 fm. From Fig. \ref{fig:9}, it can be seen
that the main contribution to the cross section comes from events where the impact
parameter is large (solid curve). Therefore, the cross section for the yield of the most
energetic light particles from nuclear reactions is governed primarily by the dynamics of
peripheral collisions, which in turn depends on the nuclear component of the interaction
potential $V_{AB}$.

\begin{figure}
\epsfxsize = 110 mm \centerline {\epsfbox{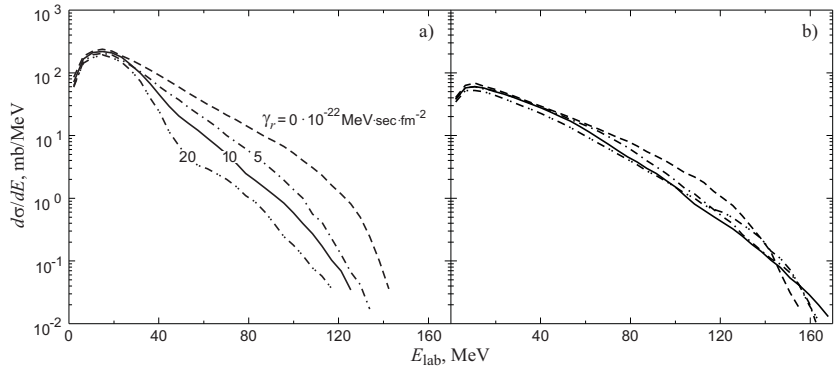}} \caption{Differential cross
sections calculated for preequilibrium neutrons emitted by (a) the projectile and (b) the
target in $^{36}Ar$(35 A MeV) + ${}^{107}Ag \to n + X$ reactions. Different curves
correspond to different values of the radial coefficient of friction $\gamma_r$. The
values of this coefficient in $10^{-22}$ MeV s fm$^{-2}$ units are indicated on each
curve in Fig. \ref{fig:10}a.} \label{fig:10}
\end{figure}

As was demonstrated above, the relative velocity of the light particle from the target
and the heavy projectile core determines the maximum angle [see Eq. (\ref{eq13})] at
which the light particle can be scattered and, hence, the maximum energy of this
particle. At beam energies $E_0$ of about the Fermi energy, the relative velocities are
so great that the angle at which the light particle is scattered is small, and so is
therefore its energy. On the other hand, it is well known from experiments that, in the
hard section of the spectrum, the energy of light particles can be as large as about
4$E_0$ to 6$E_0$. The inclusion of nuclear friction in the model being considered is a
mechanism that could ensure the reduction of the relative velocity in the scattering of
the light fragment $b$ on the projectile core $A$. As was indicated above, we introduce
only the forces of friction between the cores of the projectile and the target. The form
factor of dissipative forces is not known precisely; therefore, use was made here of a
phenomenological form factor of the Fermi type both for the radial and for the tangential
component. In \cite{r33}, it was indicated that, at the same values of the coefficient of
friction, a form factor of this type leads to identical friction between two light and
two heavy nuclei at equal distances between their surfaces. This means that, in this
case, the coefficients of friction must depend on the masses of colliding nuclei;
therefore, one cannot expect that there exist universal values of these coefficients for
any nuclear system.

For preequilibrium neutrons originating from ${}^{36}Ar$(35 A MeV) + ${}^{107}Ag \to n +
X$ reactions, we have calculated the formation cross sections at various values of the
radial coefficient of friction. In Fig. \ref{fig:10}, the results of these calculations
are presented individually for neutrons emitted by (a) the projectile and (b) the target.
From Fig. \ref{fig:10}a, it can be seen that nuclear friction exerts the strongest effect
on the cross section for the yield of projectile neutrons. The dissipation of energy
leads to a decrease in the relative velocity $v_0$ of the nuclei and, hence, to a
reduction of the maximum energy of neutrons emitted by the projectile [see Eq.
(\ref{eq10})]. For particles emitted from the target, an increase in the coefficients of
friction leads, however, to an increase in their maximum energy (by 10 to 15 MeV in the
case being considered) owing to a decrease in the energy of the relative motion of the
target neutron and the projectile and, consequently, owing to the growth of the angle of
rainbow neutron scattering in the field of the projectile [see Eqs. (\ref{eq12}),
(\ref{eq13})]. That the form of the energy spectrum of neutrons escaping from the target
shows a relatively weak dependence on dissipative nuclear forces is explained
predominantly by the peripheral character of processes leading to neutron emission.
Needless to say, an indefinite increase in the coefficient of friction would not lead to
a steady growth of the maximum energy of neutrons escaping from the target-in other

Fixing the initial configurations of the projectile and the target nucleus, the collision
energy $E_0$, and the impact parameter $\rho$, we can single out, from the entire set of
events, the following two subsets: one comprising events where the light particle is
emitted from the projectile and the other comprising events where the light particle is
emitted from the target. The initial parameters can be chosen in such a way that the
process of light-particle formation in each of these subsets would be affected
predominantly by any mechanism of those that were described in Section 3. By varying the
parameters of nuclear friction for each of these event types, one can investigate their
effect on collision dynamics and on the properties of the emitted light particle - that
is, on its energy and emission angle. Such an analysis was performed in this study for
preequilibrium neutrons originating from ${}^{36}Ar$(35 A MeV) + ${}^{107}Ag$
interaction. As was anticipated, the asymptotic energy of target neutrons accelerated by
the mean field of the projectile nucleus grows as the coefficient $\gamma_r$ is increased
up (15-20)$\times 10^{-22}$ MeV s fm$^{-2}$. A further increase in $\gamma_r$ leads to a
fast reduction of the neutron energy. The neutron emission angle in the laboratory frame
decreases with increasing radial coefficient of friction. Thus, we can state that small
friction as if focuses preequilibrium light particles from the target in the beam
direction and leads to the growth of their energy. With increasing $\gamma_r$, the energy
of a neutron emitted by the projectile nucleus decreases monotonically, while its
emission angle grows in absolute value, remaining negative. This means that, when the
velocity of the relative motion of the nuclei involved is reduced because of the effect
of dissipative forces, the projectile neutron is subjected, for a longer time, to the
effect of the attracting target field, which, as the coefficient $\gamma_r$ is increased,
distorts its trajectory ever more strongly and which, in the case of strong nuclear
friction, can even lead to neutron capture by the target. Therefore, the effect of the
growth of dissipative forces on light particles emitted from the projectile is opposite
to the effect of the analogous growth on target light particles. We also note that the
introduction of a small friction does not affect the total cross section (that is, the
total multiplicity) for the production of preequilibrium light particles, changing only
the character of the differential cross section as a function of energy and emission
angle; that is, this leads to a redistribution of emitted light particles in terms of the
coordinates $E_n$ and $\theta_n$. And only in the case of large dissipative forces does
the multiplicity of preequilibrium light particles decrease significantly. We can see
that the properties of the angular and the energy distributions of light particles are
sensitive to the form of dissipative nuclear forces; therefore, valuable information
about the magnitude of the coefficients of nuclear friction and about other parameters of
the dissipative function-in particular, about their dependence on the masses of colliding
nuclei and on energy-can be extracted from an analysis of a vast body of experimental
data on the yields of fast preequilibrium particles. By way of example, the
friction-parameter values used in calculating the cross sections for the formation of
light particles in Section 4 are given in the table. In each case, we have chosen a
radial form factor of the Fermi type; its radius $R_{fr}$ and its diffuseness parameter
$a_{fr}$ are also quoted in the table \ref{table:1}.

\begin{table}
\caption{Friction form-factor parameters} \label{table:1}
\begin{center}
\begin{tabular}{ccccc} \hline \hline
Reaction & $\gamma_r$,  & $\gamma_t$, & $R_{fr}$, & $ a_{fr}$, \\   & $10^{-22}$ MeV sec
fm$^{-2}$  & $10^{-22}$ MeV sec fm$^{-2}$ & fm & fm \\ \hline $^{132}Xe$ (44 A MeV) +
$^{197}Au$ & 45 & 0.4 & 13.1 & 0.7 \\ $^{20}Ne$ (30 A MeV) + $^{165}Ho$ & 15 & 0.1 & 9.4
& 0.7 \\ $^{36}Ar$ (35 A MeV) + $^{107}Ag$ & 10 & 0.1 & 9.3 & 0.7 \\ $^{16}O$ (20 A MeV)
+ $^{197}Au$ & 6 & 0.05 & 9.2 & 0.7 \\ \hline
\end{tabular}
\end{center}
\end{table}

\section{CONCLUSION}

In order to study intermediate-energy ($E_0 <$ 100 MeV per nucleon) heavy-ion collisions
leading to the production of fast light particles, we have proposed a classical four-body
model. Within this model, projectile and target nuclei are represented as two-particle
subsystems, each consisting of a heavy core and a light fragment (for example, a proton,
a neutron, an alpha particle, etc.). The proposed approach has been used to study in
detail the dynamics and the mechanisms of formation of preequilibrium light particles
originating from ${}^{20}Ne$(20, 30 A MeV) + ${}^{165}Ho \to n + ER$, $^{16}O$(20 A MeV)
+ ${}^{197}Au \to p +X$, ${}^{36}Ar$(35 A MeV) + ${}^{107}Ag \to n +X$, ${}^{40}Ar$(44 A
MeV) + ${}^{51}V \to p + X$, and ${}^{132}Xe$(44A MeV) + ${}^{51}V, {}^{197}Au \to p + X$
reactions treated by way of example. Our theoretical estimates agree well with
experimental data.

The role of two-body interactions in the process of light-particle formation has been
investigated in detail, and the main mechanisms of this process have been determined. It
has been shown that the high-energy component in the spectrum of neutrons and protons
from these reactions corresponds to preequilibrium particles emitted both by the
projectile and by the target nucleus. It has has been found that the yield of ultrafast
preequilibrium light particles from the target nucleus exceeds the yield of light
particles from the projectile nucleus, almost completely saturating the hardest section
of their energy spectrum. The process of target-nucleon acceleration by the projectile
mean field plays a dominant role here. Nucleon-nucleon collisions are insignificant in
this respect at energies below 50 MeV per nucleon. With increasing projectile energy, the
effect of the mean fields weakens, whereas nucleon-nucleon collisions become a dominant
process in the formation of the hard section of light-particle spectra.

It has been revealed that nuclear-friction forces strongly affect the character of the
angular and energy distributions of preequilibrium light particles. At high energies, the
introduction of dissipative forces leads, among other things, to a moderation of the
projectile, with the result that the mean fields exert a more pronounced effect on the
nucleons of colliding nuclei. The forces of friction reduce the yield of fast particles
from the projectile nucleus, but they increase the maximum energy of light particles
emitted by the target nucleus. More detailed information about the character and the
magnitude of dissipative nuclear forces would be deduced from a comprehensive analysis of
extensive experimental data on the cross sections for light-particle formation within
this approach.

The proposed new mechanism of preequilibrium-light-particle formation (acceleration of
target nucleons in the projectile mean field) is indirectly confirmed by experimental
data. In order to obtain more compelling pieces of evidence in favor of the existence of
this mechanism, one could, for example, study the spectra of neutrons or protons emitted
in collisions of two different projectile species with the same target nucleus. In doing
this, it is necessary to select the combinations of projectiles and targets in such a way
that one could separate light particles emitted by the projectile nucleus from those
emitted by the target nucleus. For this, it is required that the spectrum of light
particles originating from the projectile change insignificantly upon going over from one
projectile species to another. In this case, the change in the observed distribution of
preequilibrium light particles will be completely determined by the change in the
distribution of light particles emitted by the target.



\begin{thebibliography}{99}
\bibitem{r1} R.M. Eisberg and G.Igo, \PR{93,1954,1039}.

\bibitem{r2} R.M. Eisberg, \PR{94,1954,739}.

\bibitem{r3} T.C. Awes et al., \PR{C24,1981, 89}.

\bibitem{r4} T. C. Awes et al., \PR{C25,1982,2361}.

\bibitem{r5} T. C. Awes et al., \PL{B103,1981,417}.

\bibitem{r6} J. Kasagi, \PL{104B,1981,434}.

\bibitem{r7} B. Ludewigt et al., \PL{108B,1982,15}.

\bibitem{r8} D. Hilscher et al., \NP{A471,1987,77c}.

\bibitem{r9} E. Holub et al., \PR{C33,1986,143}.

\bibitem{r10} B. V. Jacak et al., \PR{C35,1987, 1751}.

\bibitem{r11} G. Lanzano et al., \PR{C58,1998,281}.

\bibitem{r12} Yu. E. Penionzhkevich et al., Preprint No. E7-98-282, JINR (Dubna,1998).

\bibitem{r13} D. Prindle et al., \PR{C57,1998,1305}.

\bibitem{r14} P. Sapienza et al., \NP{A630,1998,215c}.

\bibitem{r15} D. Sackett et al.,\PR{C44,1991,384}.

\bibitem{r16} R. Alba et al., \PL{B322,1994,38}.

\bibitem{r17} S. I. A. Garpman, D. Sperber, and M. Zielinska-Pfabe, \PL{90B,1980,53}.

\bibitem{r18} T. Udagawa and T. Tamura, \PRL{45,1980,1311}.

\bibitem{r19} V. E. Bunakov, V. I. Zagrebaev, and A. A. Kolozhvari, \JL{Izv.\ Akad.\ Nauk\ SSSR,\ Ser.\
Fiz., 44,1980,2331}.

\bibitem{r20} V. I. Zagrebaev and A. Yu. Kozhin, \JL{Izv.\ Akad.\ Nauk\ SSSR,\ Ser.\
Fiz., 52,1988,104}.

\bibitem{r21} V. E. Bunakov and V. I. Zagrebaev, \JL{Z.\ Phys., A333,1989,57}.

\bibitem{r22} V. I. Zagrebaev, \JL{Ann.\ Phys.\ (N.Y.), 197,1990,33}.

\bibitem{r23} V. I. Zagrebaev, in Proceedings of the First International School in Nuclear Physics,
Kiev, 1990, p. 471.

\bibitem{r24} S. Leray et al., \JL{Z.\ Phys., A320,1985,383}.

\bibitem{r25} J. P. Bondorf et al., \NP{A333,1980,285}.

\bibitem{r26} K. Mohring, W. J. Swiatecki, and M. Zielinska-Pfabe, \NP{A440,1985,89}.

\bibitem{r27} F. Sebil and B. Remaud, \JL{Z.\ Phys.,A310,1983,99}.

\bibitem{r28} J. Randrup and R. Vanderbosch, \NP{A474,1987,219}.

\bibitem{r29} V. Zagrebaev, in Proceedings of the XV Nuclear Physics Divisional Conference LEND-95,
St. Petersburg, 1995 (World Sci., Singapore, 1995), p. 457.

\bibitem{r30} V. Zagrebaev and Yu. Penionzhkevich, \JL{Prog.\ Part.\ Nucl.\ Phys., 35,1995,575}.

\bibitem{r31} C. M. Perey and F. G. Perey, \JL{At.\ Data\ Nucl.\ Data\ Tables, 13,1974,293}.

\bibitem{r32} J. Blocki,   J. Randrup, W.J.   Swiatecki and C. F. Tsang, \ANN{105,1977,427}.

\bibitem{r33} D.H.E. Gross and H.Kalinowski, \JL{Phys.\ Rep., 45,1978,175}.

\bibitem{r34} J. Knoll and R. Schaeffer, \ANN{97,1976,307}.
\end{thebibliography}
\end{document}